\documentclass{article}

\usepackage{arxiv}
\usepackage[font=sf]{caption} 

\usepackage[utf8]{inputenc} 
\usepackage[T1]{fontenc}    
\usepackage{hyperref}       
\usepackage{url}            
\usepackage{booktabs}       
\usepackage{amsfonts}       
\usepackage{nicefrac}       
\usepackage{microtype}      
\usepackage{lipsum}		
\usepackage{graphicx}
\usepackage{doi}
\usepackage{subcaption}
\usepackage{amsmath}
\usepackage{amsfonts} 
\usepackage{amssymb}
\usepackage{amsthm}
\usepackage{siunitx}
\usepackage{verbatim}
\usepackage{hyperref}
\usepackage[dvipsnames]{xcolor}
\usepackage[normalem]{ulem}

\usepackage[backend=biber]{biblatex}
\theoremstyle{definition}

\usepackage{float}
\usepackage{tikz}
\usetikzlibrary{shapes, arrows.meta, calc}
\usepackage{url}
\usepackage{titlesec}

\titleformat{\section}
  {\centering\normalfont\normalsize\bfseries\scshape}{\thesection.}{1em}{}
\titleformat{\subsection}
  {\normalfont\normalsize\bfseries}{\thesubsection.}{1em}{}
\titleformat{\subsubsection}
  {\normalfont\normalsize\itshape}{\thesubsubsection.}{1em}{}
 
\theoremstyle{definition}
\newtheorem{definition}{Definition}[section]

\newtheorem{remark}[definition]{Remark}

\title{Analyzing Access to Department of Motor Vehicles (DMV) Locations in California: A Persistent Homology Approach}

\author{
 Aylin Ambriz \\
  University of California, Berkeley\\
  Berkeley, CA 94720 \\
  \texttt{aylin.amb@berkeley.edu} \\
   \And
 Crystela Herrera \\
  University of California, Berkeley\\
  Berkeley, CA 94720 \\
  \texttt{crystelaherrera@berkeley.edu} \\
  \And
 Derek Price \\
   Tufts University\\
   Boston, MA 02155 \\
   \texttt{kmartinezb@brynmawr.edu} \\
   \And
 Robin Wilson \\
   Loyola Marymount University \\
   Los Angeles, CA 90045 \\
   \texttt{Robin.Wilson@lmu.edu} \\
   \And
 Nicholas Arosemena \\
   Brown University \\
   Providence RI 02912 \\
   \texttt{nicholas\_arosemena@brown.edu} \\
   \And
 Atilio Barreda \\
   CUNY Graduate Center \\
   New York, NY 10016 \\
   \texttt{abarreda@gradcenter.cuny.edu} \\
   \And
 Omayra Ortega \\
   Sonoma State University  \\
   Rohnert Park, CA 94928 \\
   \texttt{omayra.ortega@sonoma.edu} \\
   \And
 Daniel A. Cruz \\
   California Polytechnic State University, SLO \\
    San Luis Obispo, California 93407 \\
   \texttt{dcruz66@calpoly.edu} \\
}

\begin{document}
\maketitle

\begin{abstract}
	In 2019, the California Department of Motor Vehicles (DMV) began offering online services to alleviate in-person congestion. However, a combination of long wait times, understaffing, and systemic inequities in public service continues to make access to DMV services more cumbersome. This study aims to assess the accessibility of DMV services across California using a mathematical framework known as topological data analysis (TDA). TDA captures the shape and connectivity of service networks, revealing underserved areas. TDA has been used in other disciplines of resource coverage, including voting centers, and we build on that methodology to examine accessibility across DMV locations. We considered 167 DMV sites in California and compute pairwise travel times between each site using the Google Routes application programming interface, considering travel by car, public transportation, and walking; driving proved fastest on every pair, so the analysis that follows is an analysis of access by car. We constructed a weighted Vietoris--Rips filtration where the vertex weights represent estimated service times and edge weights represent travel times. Applying persistent homology to the resulting weighted Vietoris--Rips filtration, we obtain persistence diagrams for appointment-based and for walk-in service, and we interpret 1-dimensional features as gaps in service coverage. We localize these features geographically and compare the demographic composition of the ZIP codes they contain against the rest of the state. Coverage gaps are associated with race/ethnicity and with household income at conventional significance levels, but the effect sizes are negligible, and the direction of the income association runs opposite to what a burden-based account of DMV access would predict: travel-time coverage gaps open up in low-density, higher-income parts of the state rather than in dense urban areas.
\end{abstract}


\section{Introduction}\label{sec:intro}

\subsection{Accessibility of DMV Services}
\label{sec:motivation}
Valid identification is a critical requirement for participating in many aspects of public life, including securing employment, accessing government services, and voting in several U.S. states. In jurisdictions with voter ID laws, individuals without approved forms of ID may face barriers to exercising their democratic rights. Similarly, lack of valid ID can prevent individuals from entering the workforce or accessing essential services.

The DMV operates 176 field locations across California as of 2025, four of them Drive Test Centers. However, during our research process, the following nine locations were not in operation due to refurbishment or permanent closures that took place within the year: Banning, Capitola, Inglewood, Laguna-Hills, Oakland-Coliseum, Pacoima Driver's License Processing Center (DLPC), San Jose DLPC, and Santa-Maria. Thus our findings are in alignment with the information we had for the remaining 167 DMV sites. 

It is important to distinguish between the services offered in person as opposed to those offered online. This is especially important as not all individuals have reliable internet access or digital literacy to use online services, making physical DMV access critical. Additionally, depending on appointment availability, several people opt to be walk-in customers, so we will also be differentiating our data with respect to this criterion. 

One should note that not all DMV services are available to walk-in customers, like driving tests which require an appointment. Additionally, several DMV services, such as replacing driver's licenses or identification cards, replacing vehicle registration, reinstating driving privileges, and updating name or gender information, continue to require in-person visits, which are strongly encouraged to be scheduled by appointment. As of May 7, 2025, individuals 18 and older must have a REAL ID or another valid form of identification to board domestic flights and/ or enter specific federal buildings. This recent federal enforcement of the Real ID Act has caused an influx of congestion in DMVs across California. 

While routine services like license or registration renewals affect the general public, individuals seeking reinstatement of driving privileges or updating name and gender markers on identification cards are especially impacted by long DMV wait times at their local offices. To better understand who may be most affected by long DMV wait times, particularly those seeking to reinstate driving privileges, it's important to consider the systemic factors behind license revocations. Driving privileges are often revoked due to issues such as unpaid child support, unpaid taxes, DUIs, or excessive traffic violations. Research has shown that such penalties disproportionately impact low-income individuals and communities of color, compounding existing barriers to mobility and access to essential public services \cite{uscivilrights2017}.

In light of these issues, our study highlights the need to assess DMV accessibility not only through the lens of physical proximity and waiting times, but also through a deeper understanding of who relies on these services and why. As DMV offices face increased demand due to federal mandates like the REAL ID Act and limited online alternatives for certain populations, long wait times risk exacerbating existing inequities—particularly for individuals navigating license reinstatement or identity updates. By identifying where service bottlenecks occur and which communities are most affected, this research aims to inform more equitable policy decisions and resource allocation strategies for the future.

\subsection{Related Works} 

As mentioned previously, the use of topological data analysis (TDA) has been implemented by \cite{hickok2022} in their \textbf{Persistent Homology for Resource Coverage: A Case Study of Access to Polling Sites}, to find gaps in access to polling sites across the  major cities of Atlanta, Chicago, Jacksonville, Los Angeles, New York City, and Salt Lake City. They build a weighted Vietoris--Rips (VR) filtration on the set of polling sites, using a non-Euclidean, population-weighted travel-time dissimilarity for the edges and a site-specific quantity for the vertex weights, and they interpret persistent $1$-dimensional homology classes as holes in coverage.

Coverage questions of this kind have a longer history in applied topology, beginning with the sensor network coverage criteria of \cite{desilva2007}, where homology certifies the absence of holes in a region covered by sensors of limited range. In geographic and social applications, \cite{feng2020connecting}  survey the relevant machinery.


It is also worth positioning this working against  standard geographic information science approach to spatial accessibility, the two-step floating catchment area (2SFCA) method
\cite{luowang2003} and its other variants \cite{luoqi2009}, which assign each population unit an accessibility score by combining supply-to-demand ratios over a travel-time catchment. We use
persistent homology instead because it is "scale-free" --- it does not require committing in advance
to a catchment radius, and instead reports the range of travel times over which a gap persists. It also returns the geometry and location of a gap rather than a scalar score attached to each unit.

Our research question is: Which areas of California are underserved by DMV services and what connections are there to racial and socioeconomic demographics? TDA allows us to detect gaps, or ``holes", in service coverage that may not be evident through traditional spatial analysis alone. Although many studies rely solely on geographic distance to assess access, this approach incorporates both travel times and average wait times, providing a more complete view of how accessibility to services is experienced. 

From a topological perspective, DMV locations are modeled as vertices in a non-Euclidean metric space, where each vertex carries a weight representing the appointment or walk-in wait time at that location. An edge incident with two vertices represents the pairwise commute time between their corresponding locations via car, public transportation, or walking (in practice always by car; see \S\ref{sec:dissimilarity}). To quantify accessibility, we examine the edges and triangles that form as we increase the filtration parameter $\epsilon$. A vertex is included in our metric space once $\epsilon$ surpasses its associated wait time, and an edge forms between two vertices once $2\epsilon$ exceeds the sum of their pairwise commute time and their two wait times. Intuitively, $\epsilon$ can be thought of as a growing ball around each vertex, where an edge forms the moment two balls intersect. Following Hickok et al., we aim to identify gaps in coverage that emerge and disappear as $\epsilon$ increases.

Throughout this paper we adopt the convention of Section~2, in which a simplex
enters the complex when $d(x_i,x_j) + w_i + w_j < 2\epsilon$ holds for every
pair of its vertices. Birth and death values reported in
Table~\ref{tab:features} and Table~\ref{tab:walkin-features} are quoted on the
$2\epsilon$ scale, that is, as values of $d + w_i + w_j$ rather than of
$\epsilon$ itself. This is the natural scale for reading a death value as a
travel-plus-wait budget in minutes, but it is twice the value that a persistence
diagram plotted in $\epsilon$ would show, and we flag it here so the two are not
confused.

\section{Some Preliminaries}

We will briefly introduce pertinent mathematical concepts from TDA and specifically, Persistent Homology (PH). Please note, the following background on persistent homology is based on the given description in \cite{hickok2022}, \cite{feng2020connecting} and \cite{kemme2025persistent}. We begin with a finite collection of data points in a metric space, known as a point cloud. One way to study of the shape and structure of this point cloud is known as TDA. The primary tools used in TDA are simplicial complexes and homology groups, which are used to analyze the underlying shape and structure of the data. Applying these main tools, we arrive at persistent homology, a way to use simplicial complexes to describe features that persist in the data across multiple scales. In this context, a feature refers to a topological characteristic of the data captured by homology in a given dimension. For example 0-dimensional features are connected components, 1-dimensional features are loops or holes, and 2-dimensional features are voids or enclosed cavities. These features are equivalence classes of cycles that persist across a range of scales. We quickly define these terms below: 

For the purposes of our analysis, we say a \textbf{0-simplex} is a point in a topological space (0-Dimensional), a \textbf{1-simplex} is a line segment (1-Dimensional), a \textbf{2-simplex} is a filled triangle (2-Dimensional), and so on. We restrict to simplices up to dimension 2 (i.e., points, edges, and triangles) because our data set is 2-dimensional. To clarify, the homological features mentioned in the previous paragraph are not simplices themselves.


A \textbf{simplicial complex $K$} is a collection of simplices such that for any simplex $\kappa \in K$, every face of $\kappa$ is also in $K$. We will be considering simplicial complexes composed of up to and including 2-dimensional simplices. In our analysis, the simplicial complexes we construct contain vertices, edges, and triangles, that is, 0-, 1-, and 2-simplices. These elements allow us to compute homology in 0-Dimensional (0-D) and 1-Dimensional (1-D) homology of such simplicial complexes. The 0-D homology group captures connected components of the data, determined by how edges link together isolated points. The 1-D homology group captures "loops'' i.e. closed cycles formed by edges that are not filled in by triangles. When a triangle fills in a cycle, it effectively ``kills'' that loop from the perspective of the 1-D homology group, as the hole is no longer present. Since our complexes do not contain 3-simplices (i.e. filled tetrahedra), we do not capture higher-dimensional features like 2-D voids. Thus, our persistent homology computations are limited to 0-D and 1-D topological features, that is, space between vertices (connectedness) and gaps formed by edges (holes).

\begin{figure}[h]
  \centering
  \includegraphics[width=0.8\textwidth]{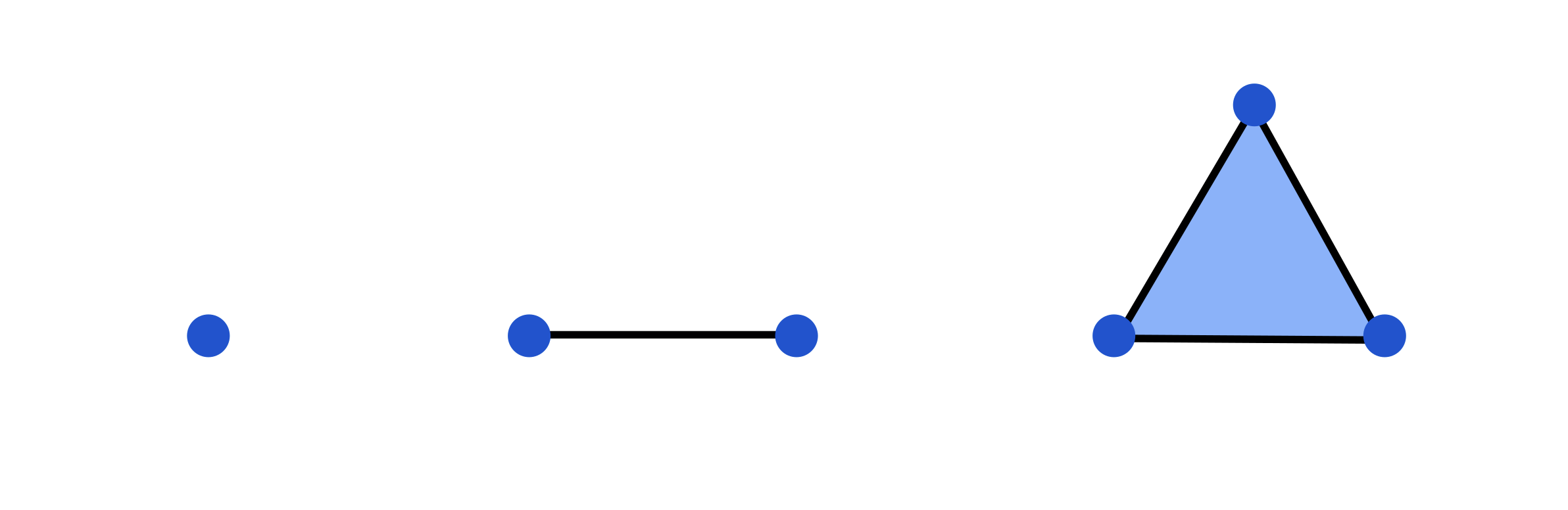}
  \caption{From left to right we have simplicial complexes composed of 0- simplices (vertex), 0- and 1- simplices (two vertices and an edge connecting them), and 0-, 1-, and 2- simplices (three vertices, three edges and a filled triangle)}
  \label{fig:simplices}
\end{figure}

\begin{figure}[h]
  \centering
  \includegraphics[width=1.0\textwidth]{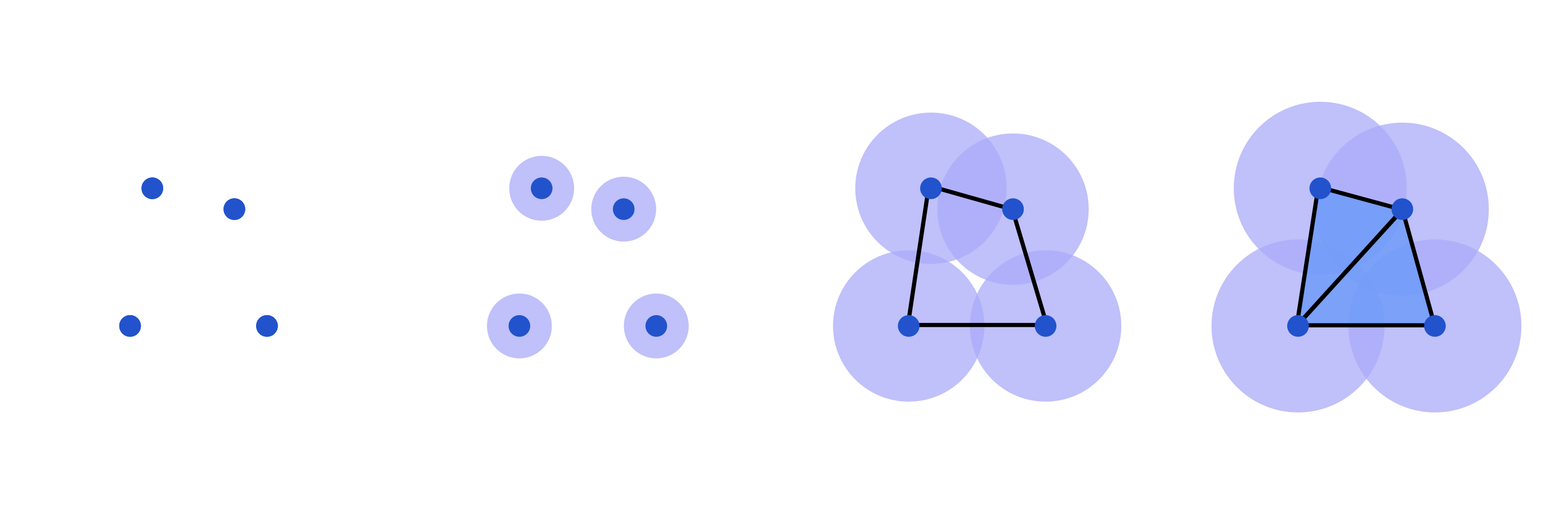}
  \caption{Depicted from left to right is a VR complex when the filtration parameter $\epsilon$ = 0, 1, 3 and 4 respectively. This nested sequence of VR complexes is an example of a VR filtration.}
  \label{fig:unweighted_vr_filtration}
\end{figure}

Given a finite set of points \( X = \{x_1, x_2, \ldots, x_n\} \)  in a metric space, the \textbf{Vietoris--Rips (VR) Complex} is a simplicial complex constructed using only the pairwise distances between points in \(X\). The vertices of the complex correspond to the elements of \(X\). A \textbf{filtration} is a nested sequence of simplicial complexes 
\( K_0 \subseteq K_1 \subseteq \cdots \subseteq K_m \), 
where each \( K_i \) includes more simplices than the previous. A VR filtration can be constructed by gradually increasing a scale parameter, \(\epsilon\), allowing one to track the evolution of topological features across different $K_i$. See Figure~\ref{fig:unweighted_vr_filtration} for an example of a VR filtration. A \textbf{filtration parameter} \(\epsilon > 0\) governs the inclusion of simplices at each stage of the filtration.

A \(k\)-simplex is included in the VR complex if the distance between every pair of its vertices \(x_i\) and \(x_j\) satisfies
\[
d(x_i, x_j) < 2\epsilon \quad \text{for all } 1 \leq j < i,
\]
where \(d(x_i, x_j)\) is the distance between points \(x_i\) and \(x_j\) as defined by the metric on \(X\). This condition corresponds to the moment when closed balls of radius \(\epsilon\) centered at each point begin to overlap pairwise.

As \(\epsilon\) increases, more simplices are added to the complex: isolated points connect via edges, loops form, and higher-dimensional simplices begin to appear, capturing the evolving topological structure of the underlying space.

Given a metric space \( X = \{x_1, x_2, \ldots, x_n\} \), a \textbf{Weighted Vietoris--Rips Complex} is a simplicial complex constructed by incorporating both pairwise distances and point-specific weights. The vertices of the complex correspond to the points in \(X\). The filtration parameter \(\epsilon\) controls when simplices are added to the complex.

\begin{figure}[h]
  \centering
  \includegraphics[width=1.0\textwidth]{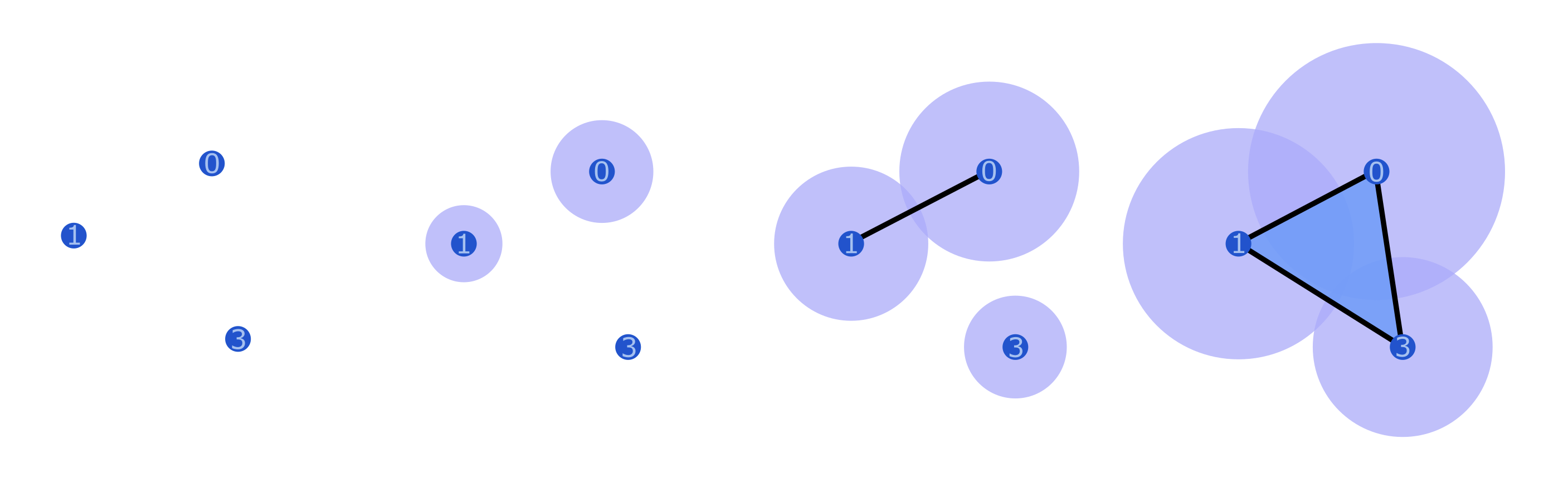}
  \caption{Depicted from left to right is a weighted VR complex when the filtration parameter $\epsilon$ = 0, 3, 6 and 9 respectively, where the weight of each node is labeled as a number. Observe that the balls growing around each node expand at different rates according to their respective weights. Nodes with larger weights require a greater filtration value before their balls intersect, thus postponing the emergence of higher-dimensional simplices.}
  \label{fig:weighted_vr_filtration}
\end{figure}

Each node \(x_i\) grows a ball of radius \(\epsilon - w_i > 0\), where \(w_i\) is the weight of node $x_i$. A \(k\)-simplex is included in the complex if and only if the distance between every pair of its vertices \(x_i\) and \(x_j\) satisfies
\[
d(x_i, x_j) + w_i + w_j < 2\epsilon \quad \text{for all } 1 \leq j < i.
\]
This condition reflects the moment at which their growing balls intersect, ensuring that simplices are added based not just on proximity, but also on the weights assigned to each vertex. As \(\epsilon\) increases, more simplices are added, leading to topological changes in the complex. See Figure~\ref{fig:weighted_vr_filtration} for an example of a weighted VR filtration. 

Persistent homology traces the birth and death of topological features such as connected components, and loops across a filtration of simplicial complexes. In our analysis the simplicial complexes we construct contain vertices, edges, and triangles, i.e. 0-, 1-, 2- simplices. These elements allow us to compute homology in dimensions 0 and 1. The 0-D homology group captures connected components of the data, determined by how edges link together isolated vertices. In 0-dimensional homology, this filtration captures the evolution of connected components. Initially, at very small $\epsilon$, each point in the cloud forms its own isolated component. As $\epsilon$ increases and edges are added between points within distance $\epsilon$, these components begin to merge. Each merge corresponds to the death of a connected component in persistent homology. Since we start with $n$ isolated vertices, and eventually end up with a single connected component, there are exactly $n - 1$ merging events --- and hence $n - 1$ finite 0-D features in the persistence diagram.

A persistence diagram is a multiset of points in the plane, where each point represents the birth and death of a topological feature during the filtration. The horizontal axis encodes the $\epsilon$ value at which the feature appears (birth), and the vertical axis records when it disappears (death), allowing us to visualize the lifespan of features across scales.

With the 1-D homology group, the weighted VR filtration captures loops that form as edges and triangles are added over time. A 1D feature is born when a loop emerges that is not yet filled in by higher-dimensional simplices (e.g. the VR complex when $\epsilon = 3$ in Figure~\ref{fig:unweighted_vr_filtration}). The loop remains topologically meaningful until it is filled by a triangle, at which point it becomes the boundary of a 2-simplex. Persistent 1D features correspond to loops that endure across a wider range of filtration values.

\section{Methodology}\label{sec:methodology}

Our methodology was inspired by the framework introduced by \cite{hickok2022} in Persistent Homology for Resource Coverage: A Case Study of Access to Polling Sites. Their study applied persistent homology to evaluate geographic access to polling sites using a weighted Vietoris–Rips(VR) filtration informed by a non-Euclidean, population-weighted distance function based on asymmetric travel times. While the underlying methodology is similar in spirit, our study adapts and extends these tools to a new domain, focusing on DMV infrastructure and its spatial and temporal dimensions of accessibility.
 

To analyze accessibility across California's DMV offices, we first compiled a comprehensive data set of 167 active DMV locations, excluding any that had been permanently or temporarily closed during our study. This data set included the name, address, ZIP code, and geographic coordinates of each office. This was our point cloud, with no metric yet defined. 

\begin{figure}[htbp]
   \centering
   \begin{subfigure}[t]{0.48\textwidth}
     \includegraphics[width=\linewidth]{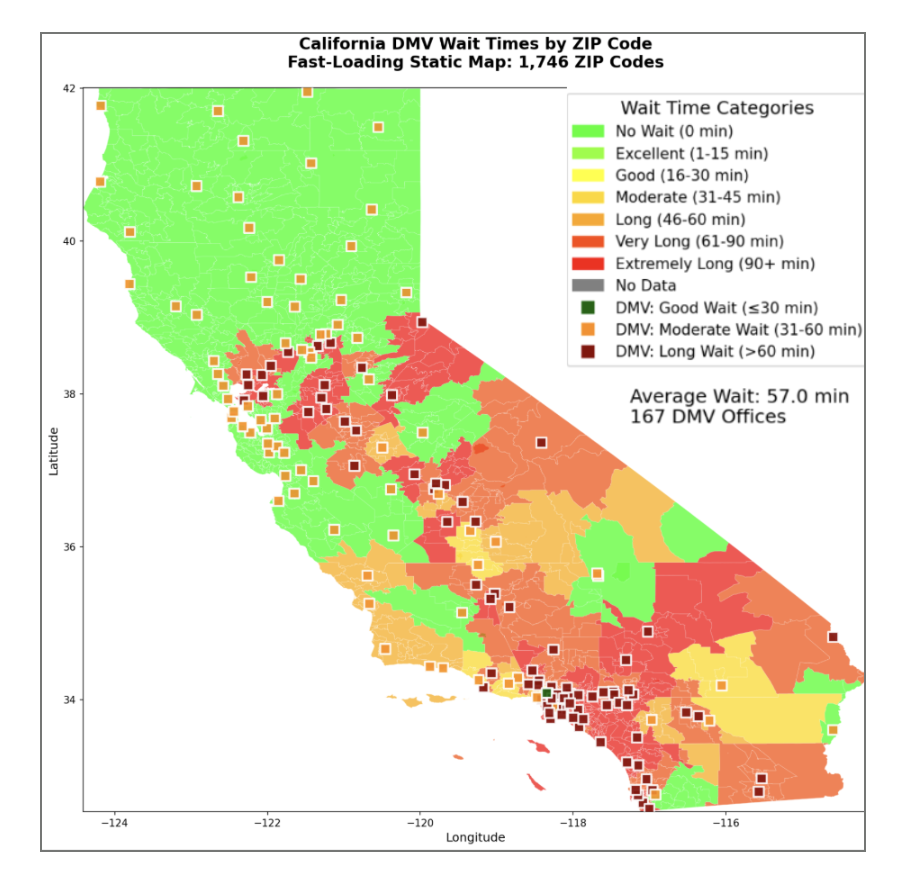}
     \caption{Walk-In average wait time: 58.4 minutes}
   \end{subfigure}
   ~
   \begin{subfigure}[t]{0.48\textwidth}
     \includegraphics[width=\linewidth]{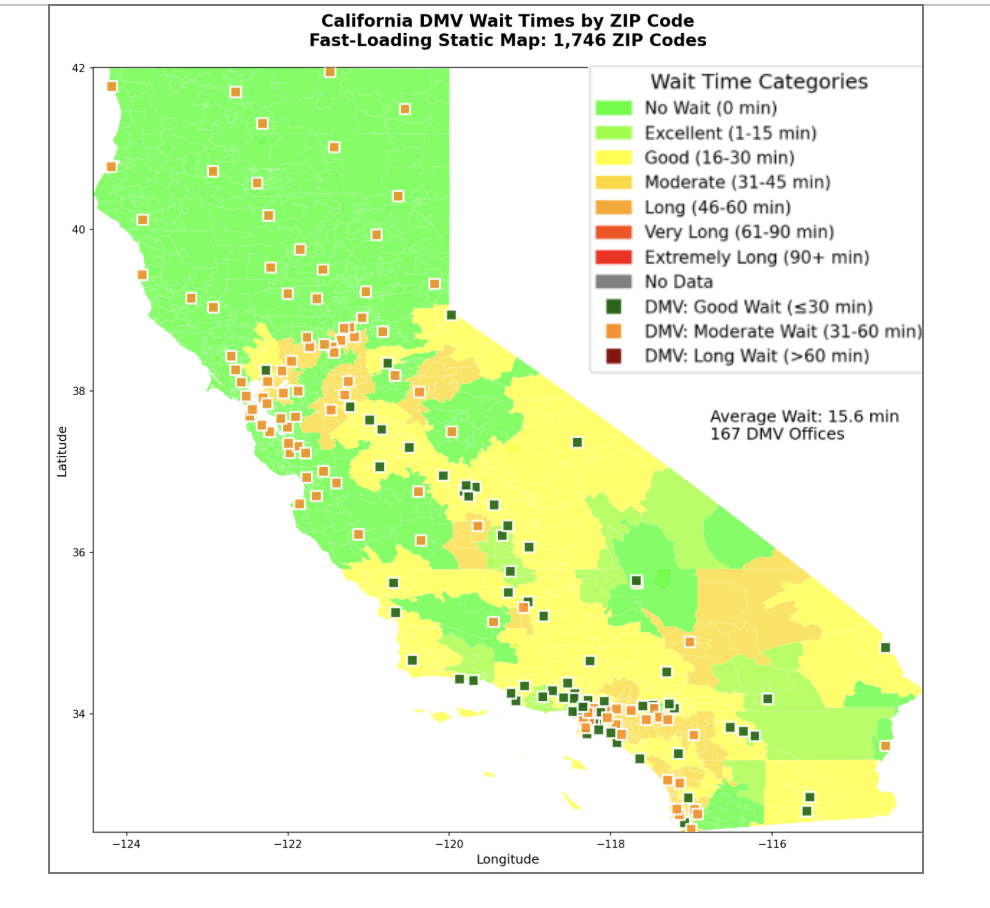}
     \caption{Appointment average wait time: 16.1 minutes}
   \end{subfigure}
   \caption{Comparison of DMV wait times by ZIP code.}
   \label{fig:waitmaps}
 \end{figure}

To fully capture DMV accessibility, we argue that travel time alone is insufficient. A complete measure of access must also account for the wait time a customer faces upon arrival. We therefore incorporate a per-location average wait time as a vertex weight in our weighted VR filtration, where heavier weights delay the formation of higher-dimensional simplices around that location, reflecting its relative inaccessibility. Since wait times differ substantially depending on whether a customer has an appointment or is a walk-in, we treat these two cases independently, constructing separate weighted VR filtrations for each and producing two distinct persistence diagrams.
Wait time data was collected for all 167 locations by web scraping DMV Wait Times, a third-party platform that aggregates historical data from the official California DMV website and accounts for time of day and day of the week to produce daily average wait times, reported as a curve over the course of the day. For each location and each day of the week, we extracted the maximum value of this curve and computed a five-day average, which we then assigned as the vertex weight for that location. Of the 167 offices, 156 were successfully geocoded during the scraping process; the remaining 11 were manually geocoded using their coordinates. We visualized the spatial distribution of these wait times using the heat map shown in Figure~\ref{fig:waitmaps}.

\subsection{A symmetric dissimilarity on the set of offices}\label{sec:dissimilarity}

In order to construct both weighted VR filtrations, we required a metric $d(x,y)$ which we constructed using the following $\tilde{d}(x,y)$ function. Using the Google Maps Routes API, we computed the travel time (in minutes) between each pair of offices for three different modes of transportation: driving, public transit, and walking. Each mode requires $167 \times 166 = 27{,}722$ ordered queries, and these values were stored as the matrices $t_{\text{car}}$ and $t_{\text{walk}}$, each $167 \times 167$, and $t_{\text{pt}}$. The transit matrix is $165 \times 165$: the API returned no transit route from or to the Bishop and Colusa offices under any query, and a further $853$ ordered pairs among the remaining offices are also missing. Because traveling from point A to point B is not the same as traveling from point B to point A, these pairwise distances were not symmetric and did not satisfy the criteria of a valid metric (we addressed this in \eqref{eq:distance}). We then defined the following function:

\begin{align}
\tilde{d}(x,y) &= \min(t_{\text{car}}(x,y), t_{\text{pt}}(x,y), t_{\text{walk}}(x,y)) \\
\tilde{d}(y,x) &= \min(t_{\text{car}}(y,x), t_{\text{pt}}(y,x), t_{\text{walk}}(y,x))
\end{align}

to represent the minimum travel time between each pair of locations across all three modes of transportation. Driving turned out to be the fastest mode on every one of the $27{,}722$ ordered pairs, so that
$\tilde{d}$ reduces exactly to $t_{\text{car}}$ and the transit and walking matrices never enter the
filtration at all. Transit, where available, took a median of $2.6$ times as
long as driving, and walking a median of $24.5$ times as long; the missing transit entries are
therefore immaterial, since transit was never the minimum on any pair where it was
available. Every travel time reported in this paper is consequently a time that a person driving a
car would experience. This is a substantive restriction rather than a technical convenience, and we return to it in
\S\ref{sec:discussion} and \S\ref{sec:future}.

Recall that the aforementioned $\tilde{d}(x,y)$ function is not a distance function as it is not symmetric, moreover, it is not a metric. Therefore, we constructed a symmetric distance function $d(x,y)$, which incorporates the populations of the ZIP codes where each DMV office is located. Incorporating the population of each ZIP code into our distance function allows us to account for the relative demand each DMV office faces. A location situated in a densely populated area is likely to serve and be accessed by far more residents than one in a sparse region. So population size serves as a proxy for the volume of people for whom that office is the most relevant point of access. ZIP-code-level population data was gathered from a California demographic data set and matched to DMV offices. The final distance between two DMV offices $x$ and $y$ was defined as:

\begin{equation}\label{eq:distance}
d(x,y) = \frac{P_z(x) \tilde{d}(x,y) + P_z(y) \tilde{d}(y,x)}{P_z(x) + P_z(y)}
\end{equation}

where $P_z(x)$ and $P_z(y)$ denote the population of the ZIP codes containing DMV offices $x$ and $y$, respectively. This function outputs a symmetric matrix with zeros along the diagonal, suitable for the construction of our weighted VR complex and filtration.

\subsection{Localizing a homology class}\label{sec:localization}

A persistent homology class is not attached to a location, so to compare
coverage gaps against demographic data we must choose a geometric
representative. We use the \emph{death simplex}: the triple of offices whose
entry into the complex fills in the loop. Drawing that triple as a triangle in
latitude and longitude gives a bounded planar region for each class, and we
classify a ZIP code as "underserved" if its polygon intersects any of these
regions.

There are a few things to note about this choice. The death simplex is not the only triangle we could have used. We could have also used the birth simplex, or the union of all simplices that make up the class, and each of these choices would give a different region. One consequence is that the corners of each triangle are DMV offices themselves, so every office sits on the edge of the region its own class defines. Under this construction, 66 of the 167 offices end up in ZIP codes that we classify as underserved. We discuss this more in \S\ref{sec:discussion}.

\subsection{Racial and socioeconomic data}\label{sec:data-demographics}

Household income data are drawn from the U.S.\ Census Bureau's American Community
Survey Subject Table S1901 (Income in the Past 12 Months) estimates
at the ZIP Code Tabulation Area (ZCTA) level, obtained from \url{data.census.gov}.
The published table reports the percentage of households in each of ten income
brackets; we converted these to household counts by multiplying by the estimated
total number of households in each ZCTA. Racial counts are reported at
the ZIP code level for seven race-alone categories, with Hispanic origin recorded
as a separate overlapping column. ZIP-code population figures $P_z$, used in
\eqref{eq:distance}, are taken from the Esri USA ZIP Code Areas layer.

Because Hispanic origin is recorded independently of race in our source, the
seven race categories partition the population and the Hispanic column cuts
across all seven; in every ZCTA the seven race counts sum exactly to the
reported total, while the Hispanic count does not. Race and Hispanic origin are
therefore not a single set of mutually exclusive groups, and where we report
them as such in Table~\ref{tab:observed_race} the construction is documented in the
Remark~\ref{rem:race-source} below.

ZIP codes are postal delivery constructs rather than areal units and do not
coincide exactly with ZCTAs; we shall treat the two as interchangeable throughout,
which introduces error at boundaries, and we say ZIP code in what follows. Three
lists are  thus involved: the coverage mapping, the income file, and the race file, and
the draft analyses used all three separately. We then restrict throughout to
their intersection within the range 90001--96162, which is $1{,}705$ ZIP codes,
of which $467$ are classified as underserved under the appointment filtration.
All tables below are computed on this set, so that the baseline shares can be
compared across tables.

\section{Results}\label{sec:result}

We utilized the GUDHI library \cite{gudhi} in Python to conduct the persistence homology process, visualized by persistence diagrams as it provides a robust suite of tools essential for performing TDA. We implicitly constructed two weighted VR filtrations through this process, one for walk-in wait times  and the other for appointment wait time. 


Since walk-in and appointment-based access to DMV services represent two independent modes of service, we analyzed them separately. Below, we present the persistence diagrams for both walk-in and appointment-based DMV service accessibility.

 \begin{figure}[H]
     \centering
     \includegraphics[width=0.6\textwidth]{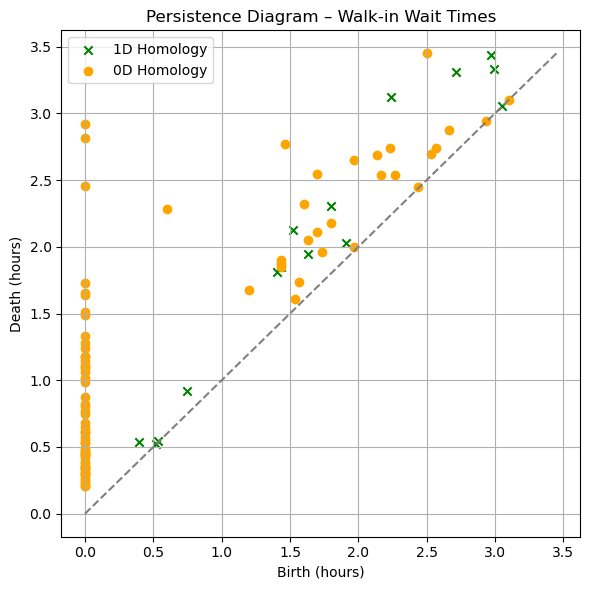}
     \caption{Persistence Diagram – Walk-in Wait Times}
     \label{fig:pd-walkin}
 \end{figure}
Figure~\ref{fig:pd-walkin} shows the persistence diagram of the walk-in filtration. A striking
feature is the dense cluster of $0$-dimensional points near the vertical axis: these are the offices
with very short walk-in wait times, which enter the filtration immediately and merge quickly with
their neighbors. The remaining $0$-dimensional points, further to the right, are offices with long
walk-in waits; they enter late and merge gradually, which is consistent with spatial separation or
with clusters of similarly congested sites. The $1$-dimensional features appear at later filtration
values, as one would expect for classes that require a connected structure to form at all, and the
more persistent among them indicate configurations where a set of low-wait offices
surrounds a region that is not well served.

The two filtrations differ less than one might expect. The median value of $d$ over all pairs of
offices is $250.8$ minutes, against a median vertex weight of $19.0$ minutes for appointments and
$63.0$ minutes for walk-ins; in both filtrations the geometry of the state contributes
more than the wait times do. The walk-in filtration yields $15$ finite $1$-dimensional classes to
the appointment filtration's $19$, and nine of the fifteen have the same death simplex in both. We thus
base the demographic analysis on the appointment filtration not because the walk-in structure is
moot, but because several DMV services, such as obtaining a reduced-fee or no-fee
identification card, can only be completed with an appointment. The walk-in classes are listed in
Table~\ref{tab:walkin-features}, and \S\ref{sec:results-sensitivity} repeats the demographic
comparison under the walk-in filtration.

\begin{table}[htbp]
\centering
\small
\begin{tabular}{rrrrlc}
\toprule
\# & Birth & Death & Persistence & Death simplex & In appt.\\
\midrule
1 & 134.5 & 187.5 & 53.0 & Weaverville, Redding, Eureka & yes \\
2 & 162.8 & 198.7 & 35.9 & Taft, Santa Barbara, Goleta & no \\
3 & 91.5 & 127.4 & 35.9 & Susanville, Red Bluff, Chico & yes \\
4 & 108.1 & 138.1 & 30.0 & San Andreas, Pleasanton, Mariposa & no \\
5 & 178.1 & 206.3 & 28.2 & Rancho San Diego IBC, Indio, Fontana CDTC & no \\
6 & 86.0 & 110.8 & 24.8 & Novato, Lakeport, Colusa & yes \\
7 & 84.1 & 108.9 & 24.7 & Mendota, Hollister, Coalinga & no \\
8 & 179.8 & 199.8 & 20.0 & Yreka, Weaverville, Crescent City & yes \\
9 & 98.0 & 116.9 & 18.8 & Truckee, Quincy, Oroville & yes \\
10 & 44.9 & 55.1 & 10.1 & Yuba City, Oroville, Colusa & yes \\
11 & 23.5 & 32.3 & 8.8 & San Mateo, Oakland, Fremont & yes \\
12 & 114.7 & 121.6 & 6.9 & Tulare, Delano, Coalinga & no \\
13 & 31.9 & 32.5 & 0.6 & Watsonville, Seaside, Salinas & yes \\
14 & 183.0 & 183.4 & 0.4 & Rancho San Diego IBC, Hemet, Fontana CDTC & no \\
15 & 31.2 & 31.3 & 0.1 & Santa Clara, San Jose, Redwood City & yes \\
\bottomrule
\end{tabular}
\caption{The $15$ one-dimensional homology classes of the walk-in filtration, ordered by
persistence, on the same scale as Table~\ref{tab:features}. The final column records whether the
same death simplex also occurs in the appointment filtration. The six that do not are concentrated
in the south and in the interior.}
\label{tab:walkin-features}
\end{table}

 \begin{figure}[H]
     \centering
     \includegraphics[width=0.6\textwidth]{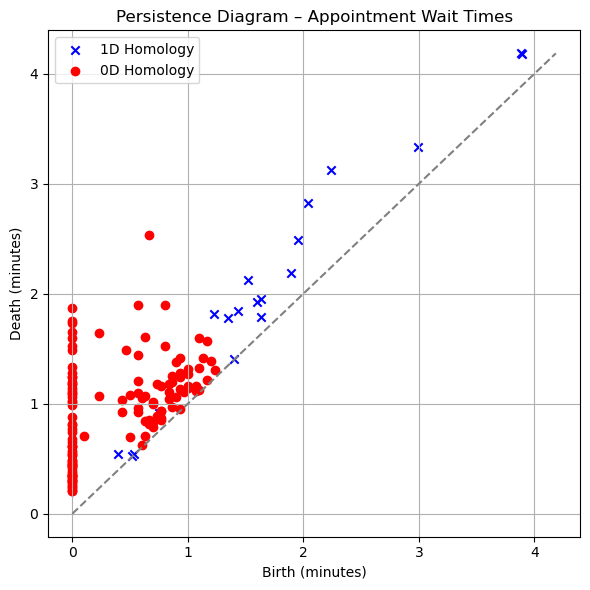}
     \caption{Persistence Diagram – Appointment Wait Times}
     \label{fig:pd-appt}
 \end{figure}

The persistence diagram (Figure~\ref{fig:pd-appt}) for appointment waiting time illustrates the topological features that emerge from applying a weighted VR filtration to our point cloud representing 167 DMV locations throughout California. In this context, the filtration is weighted by wait times, meaning that DMV locations with shorter or zero wait times are incorporated into the complex earlier than those with longer wait times. The diagram plots two types of topological features: 0-D homology (connected components) shown as red circles, and 1-D homology (loops) shown as blue crosses. Each point represents the birth and death of a feature, with more persistent features lying farther from the diagonal, where persistence is defined to be the difference of death from birth i.e. death-birth.

A notable aspect of this diagram is the clustering of the red points (0-D features) along the y-axis. These correspond to DMV locations with zero wait time, which are incorporated into the topological structure almost immediately, hence their birth times are near zero. Their deaths occur slightly later as they merge with the other components, reflecting how quickly these zero wait time DMVs connect into larger clusters in the weighted filtration. In contrast, DMVs with longer wait times appear further to the right on the x-axis, indicating that they were added to the filtration later due to their higher weights (longer wait time).

The blue crosses representing 1-D homology indicate the formation and disappearance of loops. These loops are created when the weighted structure becomes sufficiently connected to form cycles. Persistent 1-D features, those that lie far from the diagonal, suggest meaningful topological structures, such as spatial clusters of DMVs where long wait-time locations are encircled by faster ones, creating gaps or ring-like features in the point cloud. Overall, the diagram reveals that low wait time DMVs are rapidly integrated, whereas high wait time DMVs influence the formation of more delayed topological features.

After generating both persistence diagrams, we chose to focus on appointment wait times when doing further analysis. We note that several DMV services such as obtaining a Reduced/No fee identification card, can only be completed with an appointment. Therefore, we mapped the corresponding topological features from the appointment wait times persistence diagram onto a geographic visualization of California to better understand the spatial distribution of homologies across the state. 

\subsection{Geographic localization of the coverage gaps}\label{sec:results-map}
 
\begin{figure}
    \centering
    \includegraphics[width=\textwidth]{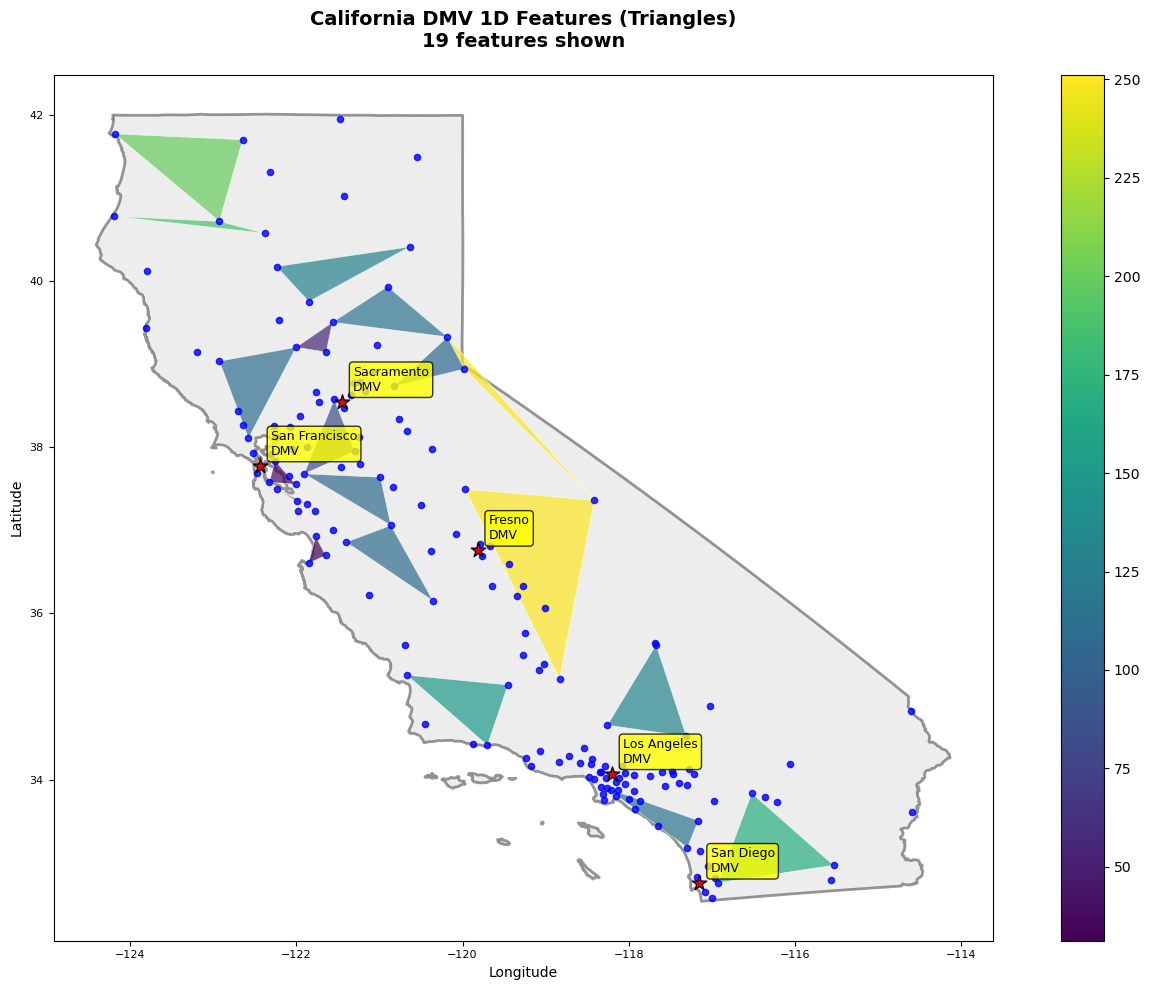}
    \caption{The death simplices of the $19$ one-dimensional classes of the appointment filtration,
    drawn over the map of California.}
    \label{fig:triangle_features}
\end{figure}

The appointment filtration has $19$ finite $1$-dimensional classes. We mapped the death simplex of
each onto California as described in \S\ref{sec:localization}; the result is
Figure~\ref{fig:triangle_features}, and the features themselves are listed in
Table~\ref{tab:features}.
 
\begin{table}[htbp]
\centering
\small
\begin{tabular}{rrrrl}
\toprule
\# & Birth & Death & Persistence & Death simplex \\
\midrule
1 & 134.5 & 187.5 & 53.0 & Weaverville, Redding, Eureka \\
2 & 122.6 & 169.7 & 47.0 & Rancho San Diego IBC, Palm Springs, Brawley \\
3 & 91.5 & 127.4 & 35.9 & Susanville, Red Bluff, Chico \\
4 & 73.3 & 108.9 & 35.5 & Los Banos, Hollister, Coalinga \\
5 & 117.2 & 149.5 & 32.3 & Taft, Santa Barbara, San Luis Obispo \\
6 & 80.7 & 106.7 & 26.0 & Pleasanton, Modesto, Los Banos \\
7 & 86.0 & 110.8 & 24.8 & Novato, Lakeport, Colusa \\
8 & 179.8 & 199.8 & 20.0 & Yreka, Weaverville, Crescent City \\
9 & 96.0 & 115.6 & 19.6 & Temecula, Oceanside, Gardena Comm.\ DTC \\
10 & 98.0 & 116.9 & 18.8 & Truckee, Quincy, Oroville \\
11 & 233.1 & 251.1 & 18.0 & Truckee, South Lake Tahoe, Bishop \\
12 & 113.5 & 131.0 & 17.5 & Victorville, Lancaster, Lake Isabella \\
13 & 233.5 & 251.0 & 17.4 & Mariposa, Bishop, Arvin \\
14 & 44.9 & 55.1 & 10.1 & Yuba City, Oroville, Colusa \\
15 & 98.2 & 107.1 & 8.9 & Truckee, South Lake Tahoe, Placerville \\
16 & 23.5 & 32.3 & 8.8 & San Mateo, Oakland, Fremont \\
17 & 31.9 & 32.5 & 0.6 & Watsonville, Seaside, Salinas \\
18 & 83.7 & 84.2 & 0.4 & W.\ Sacramento Comm.\ DTC, Stockton, Pleasanton \\
19 & 31.2 & 31.3 & 0.1 & Santa Clara, San Jose, Redwood City \\
\bottomrule
\end{tabular}
\caption{The $19$ one-dimensional homology classes of the appointment filtration, ordered by
persistence. Birth, death, and persistence are filtration values in minutes, quoted on the
$2\epsilon$ scale as described in \S\ref{sec:intro}. The death simplex is
the triple of offices whose entry into the complex fills in the loop.}
\label{tab:features}
\end{table}
 
We treat all $19$ classes as coverage gaps rather than imposing a persistence threshold, since we
have no principled basis for choosing one; Table~\ref{tab:features} allows the reader to apply any
threshold they prefer. Note that persistence and death time rank the features quite differently. The
death value is the filtration scale at which a gap finally closes, and for a coverage problem it
carries a direct interpretation (roughly, how far one must be willing to travel before the region
is enclosed) whereas persistence measures how long the gap survives once it has opened. Feature
$13$, for instance, is thirteenth by persistence but second by death time.

\subsection{Race and ethnicity}\label{sec:results-race}
 
ZIP codes that were either contained within or intersected by any of the $19$ triangles were
classified as underserved, and those not intersected by any triangle were classified as not underserved.
To assess whether there is a statistically significant relationship between ZIP code coverage status
and demographics, we conducted a $\chi^2$ test of independence. Our hypotheses are:
 
$H_0$: There is no difference between the demographics of those living in the gaps in coverage
versus those who do not.
 
$H_1$: There is a difference between the demographics of those living in the gaps in coverage versus
those who do not.
 
The racial and ethnic groups represented in the table are White (Not Hispanic), Black (Not
Hispanic), Asian (Not Hispanic), American Indian and Native Hawaiian (Not Hispanic), Hispanic (Any
Race), and Two or More Races or Other Race (Not Hispanic).

\begin{table}[htbp]
\centering
\small
\begin{tabular}{lrrr}
\toprule
Group & Not underserved & Underserved & \% underserved \\
\midrule
White       & 9{,}722{,}204  & 3{,}834{,}847 & 28.29 \\
Black       & 1{,}574{,}328  & \phantom{0}495{,}899 & 23.95 \\
Asian       & 4{,}349{,}937  & 1{,}552{,}403 & 26.30 \\
Indigenous  & \phantom{0,}161{,}568 & \phantom{00}78{,}087 & 32.58 \\
Hispanic    & 11{,}735{,}825 & 3{,}881{,}207 & 24.85 \\
Other       & 1{,}322{,}620  & \phantom{0}490{,}510 & 27.05 \\
\midrule
Total       & 28{,}866{,}482 & 10{,}332{,}953 & 26.36 \\
\bottomrule
\end{tabular}
\caption{Racial and ethnic composition of California ZIP code regions by coverage status, with the
share of each group living in an underserved ZIP code.}
\label{tab:observed_race}
\end{table}

The $\chi^2$ test rejects independence, $\chi^2(5) = 55{,}627.70$, $p < .001$, but with
$N = 39{,}199{,}435$ the effect size is negligible: Cram\'er's $V = 0.038$. The final column of
Table~\ref{tab:observed_race} makes the practical magnitude visible. Against a statewide baseline of
$26.4\%$, the share of each group living in an underserved ZIP code ranges from $23.9\%$ for Black
residents to $32.6\%$ for Indigenous residents. White, Indigenous, and Other residents are somewhat
more likely than average to live in an underserved ZIP code; Black and Hispanic residents are
somewhat less likely. The differences are real but small, and they run opposite to what the
motivating literature on the burden of DMV access would predict.

\begin{remark}\label{rem:race-source}
Table~\ref{tab:observed_race} uses six categories that do not overlap, which requires knowing how many people of each race category are not Hispanic. Our race source does not report on this. In that file, the seven race counts add up exactly to the total for each ZIP code, and Hispanic origin is reported ina separate column that overlaps all seven columns.
Table~\ref{tab:observed_race_alone} therefore reports the same comparison in the
form the source actually supports: seven race-alone categories tested together,
with Hispanic origin tested separately. Replacing the source
with ACS Table B03002 (Hispanic or Latino Origin by Race) would supply the
exclusive categories directly and would be the intended fix.
\end{remark}

\begin{table}[htbp]
\centering
\small
\begin{tabular}{lrrr}
\toprule
Group & Not underserved & Underserved & \% underserved \\
\midrule
White                                      & 12{,}455{,}111 & 4{,}754{,}179 & 27.63 \\
Black or African American                  & \phantom{0}1{,}645{,}595 & \phantom{0,}518{,}937 & 23.97 \\
Asian                                      & \phantom{0}4{,}413{,}329 & 1{,}571{,}592 & 26.26 \\
American Indian and Alaska Native          & \phantom{00,}323{,}777 & \phantom{0,}120{,}592 & 27.14 \\
Native Hawaiian and Other Pacific Islander & \phantom{00,}100{,}201 & \phantom{00,}47{,}299 & 32.07 \\
Two or More Races                          & \phantom{0}4{,}733{,}635 & 1{,}665{,}835 & 26.03 \\
Other Race                                 & \phantom{0}5{,}164{,}926 & 1{,}648{,}159 & 24.19 \\
\midrule
Total                                      & 28{,}836{,}574 & 10{,}326{,}593 & 26.37 \\
\midrule
Hispanic (any race)                        & 11{,}729{,}414 & 3{,}879{,}763 & 24.86 \\
Not Hispanic                               & 17{,}107{,}160 & 6{,}446{,}830 & 27.37 \\
\bottomrule
\end{tabular}
\caption{The same comparison using the seven race-alone categories the source supports, on the
common universe of $1{,}705$ ZIP codes. Hispanic origin is reported separately because it overlaps all seven
categories. For the seven categories, $\chi^2(6) = 40{,}050.29$, $p < .001$, $V = 0.032$; for
Hispanic origin, $\chi^2(1) = 30{,}578.83$, $p < .001$, $V = 0.028$. The ordering matches
Table~\ref{tab:observed_race}: White above the baseline, Black and Hispanic below it, with the
largest share among Native Hawaiian and Other Pacific Islander residents --- a group of $147{,}500$
people statewide, so that this last figure should be read with corresponding caution.}
\label{tab:observed_race_alone}
\end{table}
 
\subsection{Household income}\label{sec:results-income}
 
We repeated the analysis with household counts by income bracket. Our hypotheses in this case are:
 
$H_0$: There is no difference between the household income of those living in the gaps in coverage
versus those who do not.
 
$H_1$: There is a difference between the household income of those living in the gaps in coverage
versus those who do not.

\begin{table}[htbp]
\centering
\small
\begin{tabular}{lrrrr}
\toprule
Income bracket & Underserved & Not underserved & \% underserved & Pearson residual \\
\midrule
$<$\$10K       & \phantom{0}155{,}712 & \phantom{0,}458{,}620 & 25.35 & $-12.26$ \\
\$10K--\$15K   & \phantom{0}125{,}618 & \phantom{0,}381{,}156 & 24.79 & $-18.92$ \\
\$15K--\$25K   & \phantom{0}235{,}382 & \phantom{0,}673{,}374 & 25.90 & $-4.57$ \\
\$25K--\$35K   & \phantom{0}239{,}090 & \phantom{0,}689{,}941 & 25.74 & $-7.75$ \\
\$35K--\$50K   & \phantom{0}342{,}618 & \phantom{0,}972{,}368 & 26.05 & $-2.06$ \\
\$50K--\$75K   & \phantom{0}515{,}693 & 1{,}490{,}422 & 25.71 & $-12.20$ \\
\$75K--\$100K  & \phantom{0}429{,}479 & 1{,}185{,}959 & 26.59 & $10.92$ \\
\$100K--\$150K & \phantom{0}585{,}680 & 1{,}648{,}228 & 26.22 & $2.08$ \\
\$150K--\$200K & \phantom{0}327{,}968 & \phantom{0,}898{,}859 & 26.73 & $12.70$ \\
\$200K+        & \phantom{0}466{,}623 & 1{,}272{,}060 & 26.84 & $17.82$ \\
\midrule
Total          & 3{,}423{,}863 & 9{,}670{,}987 & 26.15 & \\
\bottomrule
\end{tabular}
\caption{Household income composition of California ZIP code regions by coverage status, with the
share of households in each bracket living in an underserved ZIP code and the Pearson residual of
the underserved cell.}
\label{tab:observed_income}
\end{table}

Again the test rejects independence, $\chi^2(9) = 1{,}821$, $p < 2.2 \times 10^{-16}$, and again the
effect size is negligible, $V = 0.012$. A residual near $0$ would indicate the cell behaves as expected;
a large positive residual indicates more households than expected and a large negative one fewer,
with absolute values above $2$ conventionally taken as noteworthy. The pattern is monotone in the
aggregate: underserved ZIP codes contain fewer households than expected in every bracket below
\$75K and more than expected in every bracket above it. This means the share of households
in an underserved ZIP code rises steadily as one moves up the brackets, from $24.8\%$ in the \$10K--\$15K bracket to $26.8\%$ in the
\$200K$+$ bracket, a spread of about two percentage points around a baseline of $26.1\%$.
This steady rise across brackets is, however, the one result in the paper that does not survive a change in the wait
time used to build the filtration; see \S\ref{sec:results-sensitivity}.

\subsection{Sensitivity to the choice of wait time}\label{sec:results-sensitivity}

The demographic comparisons above are built on the appointment filtration.
Since appointment and walk-in waits differ substantially (median
$19.0$ against $63.0$ minutes) it is reasonable to ask how much of the
pattern is a consequence of this choice rather than of the geometry of the office network. We
therefore repeated the entire localization and cross-tabulation using the death
simplices of the walk-in filtration listed in Table~\ref{tab:walkin-features},
holding the dissimilarity $d$ and the intersection rule of
\S\ref{sec:localization} fixed and changing only the vertex weights. The walk-in
filtration marks $331$ ZIP codes as underserved against the appointment
filtration's $472$, with $225$ in common.

The two associations behave  differently under this substitution. The
income pattern does not survive it. Under the appointment filtration the share
of households in an underserved ZIP code rises steadily from one bracket to the next, from $25.3\%$ at the bottom to $26.8\%$ at the top, a total spread of $1.49$
percentage points. Under the walk-in filtration the spread falls to $0.35$
points and the rise is no longer steady: the share peaks in the \$50K--\$75K
bracket and falls again above it. Whatever the income association is measuring, it
is not stable against a change in the wait time used, and we do not regard it as
a finding.

The race association does survive, and actually strengthens. Cram\'er's $V$ rises from
$0.032$ to $0.041$, and the ordering of groups is preserved: White and Native
Hawaiian and Other Pacific Islander residents above the baseline share, Black
and Asian residents below it, in both filtrations. This is consistent with the
reading advanced in \S\ref{sec:discussion}, on which the race pattern reflects
the spatial distribution of the office network --- which is common to both
filtrations --- rather than anything about wait times.

It is important to note that neither of these comparisons is a statistical test in the usual sense since we do not work with a sample. We have every Califronia ZIP code in the range 90001--9612, and each one is classified as underserved or not underserved depending on whether it touches one of the ninteen triangles, which is determined by where it lies on the map. No ZIP code was randomly selected into either group and there is no larger set of ZIP codes that ours represent. For this reason we treat the $\chi^2$ statistics as a description of how the two groups differ, and ask the viewer to read the effect sizes and percentages rather than the $p$-values. We say more about this in \S\ref{sec:lim-stats}.

\begin{table}[htbp]
\centering
\small
\begin{tabular}{lrr}
\toprule
 & Appointment & Walk-in \\
\midrule
Underserved ZIP codes (of $1{,}705$)        & 472    & 331    \\
Baseline share of households underserved    & 26.15\% & 19.18\% \\
Income spread, lowest to highest bracket    & $+1.49$ pp & $+0.35$ pp \\
Share rises steadily across brackets        & yes    & no     \\
Cram\'er's $V$, income                      & 0.012  & 0.012  \\
Cram\'er's $V$, race                        & 0.032  & 0.041  \\
\midrule
\multicolumn{3}{l}{\emph{Share of each group in an underserved ZIP code}} \\
White                                       & 27.61\% & 20.55\% \\
Black or African American                   & 23.95\% & 17.54\% \\
Asian                                       & 26.27\% & 16.20\% \\
American Indian and Alaska Native           & 27.13\% & 21.52\% \\
Native Hawaiian and Other Pacific Islander  & 32.05\% & 26.27\% \\
Two or More Races                           & 26.03\% & 19.78\% \\
Other Race                                  & 24.19\% & 20.31\% \\
Hispanic (any race)                         & 24.85\% & 19.60\% \\
\bottomrule
\end{tabular}
\caption{The demographic comparison repeated under both filtrations. Only the vertex weights change;
the dissimilarity and the localization rule are held fixed. Shares are not comparable in level
across columns, since the walk-in filtration marks fewer ZIP codes as underserved; the comparison of
interest is each column against its own baseline.}
\label{tab:sensitivity}
\end{table}

\section{Discussion}\label{sec:discussion}

Both associations point the same way: the ZIP codes our filtration marks as underserved skew
slightly whiter, slightly more Indigenous, and slightly higher-income than the state as a whole.
This is the opposite of what \S\ref{sec:motivation} would lead one to expect, and it is worth being
precise about why this might be true.

Our construction measures coverage on the resource side. A $1$-cycle is long-lived when a ring of
offices is mutually far apart in travel time, and the regions this picks out in California are
low-density. This is visible directly in Table~\ref{tab:features}: the leading features join
Weaverville, Redding and Eureka in the far north; Susanville, Red Bluff and Chico; Los Banos,
Hollister and Coalinga in the interior Coast Ranges. Such regions are, on average, whiter and (once one conditions on being far from a city) not especially poor, while the dense urban areas
where low-income and Hispanic households are concentrated are precisely the areas where many offices are
within a short drive. The three shortest-lived features in Table~\ref{tab:features} --- Santa
Clara--San Jose--Redwood City, West Sacramento--Stockton--Pleasanton, and
Watsonville--Seaside--Salinas --- each join three offices that already sit close to one another, so
the loops they enclose are small and close almost as soon as they open. A travel-time coverage gap is therefore a
statement about the geometry of the office network, not about the burden that a DMV visit imposes on
a household. Burden would likely depend on car ownership, on hourly versus salaried work, on the ability to take
a weekday morning off, and on how many trips a given service requires; none of which enters our
filtration.

The restriction to driving matters here in particular. Because driving was fastest on every pair,
every travel time in our filtration is a time that a person with a car would experience. A household
without one faces the transit times instead, which run a median of $2.6$ times longer where transit
exists at all, and on some pairs no transit route exists. So when we call a region a coverage gap,
what we have shown is that it is a gap for drivers. Whether it is also a gap for a household without
a car is a question the construction cannot answer --- and those are the households
\S\ref{sec:motivation} takes as its subject.

The evidence for this reading is internal to our own construction: $66$ of the $167$
offices sit in ZIP codes it labels underserved. This follows from
\S\ref{sec:localization}, since the vertices of a death simplex are offices, but it is not just a 
bookkeeping artifact. It means that around forty percent of California's DMV offices lie inside a
region that we call a coverage gap, which is only coherent if what the method identifies is a
configuration of offices rather than a population that lacks service.
 
A second point concerns aggregation method. Statewide cross-tabulation in our case dilutes local structure, and the
most conspicuous features may be telling a different story than the average one. Visalia, for
instance, lies inside the death simplex of feature $13$, spanned by the offices at Mariposa, Bishop,
and Arvin; it is $62.3\%$ Hispanic with a median household income of \$42{,}800
\cite{statisticalatlas2025}, a that is unlike the aggregate picture of
\S\ref{sec:results-race}. Visalia also has a DMV office, at 1711 East Main Street, and that office
is one of the $167$ points in our cloud. A resident of Visalia does not face a long drive to reach
service; the office is in town. What feature $13$ records is that Mariposa, Bishop, and Arvin are
far from one another, which is a fact about the sparseness of the southern Sierra. 

This feature is the second-latest to die of the nineteen, at $251.0$ minutes, but ranks only
thirteenth by persistence. That combination is exactly the profile of a large, sparsely served
region: the gap does not survive an especially wide band of filtration values, but it does not close
until the travel-time scale is very large. If a coverage gap is meant to identify places where
residents must travel a long way to reach any office, death time would be the more natural statistic to
rank by, and it selects a different set of regions than persistence does.
 
If features localized in this way do carry information that the statewide aggregate washes out, then
a ranking of features --- by death time, by persistence, or by both --- rather than membership
in a triangle is the quantity that one should also consider in the demographic comparison. That is a concrete
methodological suggestion this data set is well placed to test.

To measure the burden a DMV visit would actually place on people, we could consider a point cloud made of population rather than offices. One construction could be to place the vertices at population
"centers" and lets the vertex weight be the travel time to the nearest office, so that a
long-lived class would then mark a region of people who are all far from service as opposed to a region of
offices that are all far from each other. Under such a filtration this would avoid the problems described above, since an office is not a vertex and so cannot lie inside its own gap. This construction
is a different filtration and we have not yet built it.

\section{Conclusion}\label{sec:conclusion}

\subsection{Summary}\label{sec:summary}
In this study, we used topological data analysis—specifically persistent homology—to investigate the accessibility of DMV services across California. By constructing a weighted Vietoris–Rips filtration informed by travel times and wait times, we modeled 167 DMV locations as a point cloud and analyzed the topological structure of their spatial and temporal accessibility. We visualized service coverage through persistence diagrams and identified 1-dimensional features representing potential coverage holes. These features were mapped geographically to identify underserved regions, which we then analyzed in relation to racial and socioeconomic demographics. A chi-square test revealed significant associations between underserved status and both race/ethnicity and household income, but with negligible effect sizes and in a direction opposite to what a burden-based account
might predict. We argue that this is a consequence of measuring coverage on the resource side rather than on
the population side. Repeating the comparison under the walk-in filtration showed that the income
association is an artifact of the appointment weights and does not survive, while the race
association does and actually strengthens.
Our approach provides a novel mathematical framework for evaluating public service coverage and highlights the importance of considering both spatial and demographic factors in equitable infrastructure planning.

\section{Limitations}\label{sec:limitations}
There were several data limitations. To begin, the demographic data used was drawn from the U.S. Census Bureau’s estimates at the ZIP code level, which provides only approximated values with associated margins of error. Additionally, estimated DMV wait times were based on a five-day average collected from publicly available scheduling data; however, not all DMV locations operate five days a week, which could skew the average. 
\subsection{Modeling}
\label{lim-model}
The point cloud consists of offices rather than of people, so our coverage gaps are a statement about the office networks. The fastest mode was driving for every pair, so the analysis we completed is effectively a car-access analysis.  We also measure travel time and wait time both in mintues and add them together, but they are not really the same quantity; an hour spent driving for example is not an hour spent waiting for a service, and someone deciding whether to make the trip could weight the two differently. Our filtration treats a minute of each as equal, and we have not yet checked what would change if one were "scaked up" against the other. Finally, we had to pick one triangle to stand in for each coverage gap, and the death simplex is one of many reasonable choices.

We have, however, tested sensitivity to the level of the vertex weights, by rerunning the entire
analysis under the walk-in filtration (\S\ref{sec:results-sensitivity}); this is a weaker check than
varying the relative scaling of the two quantities continuously, but it is the substitution with the
clearest substantive interpretation.
\subsection{Statistical inference}\label{sec:lim-stats}
 
The $\chi^2$ tests treat individuals and households as independent observations. They are not:
residents of the same ZIP code share an underserved status by construction, and neighboring ZIP
codes are spatially autocorrelated. The effective sample size is therefore closer to the number of
ZIP codes than to the number of residents, the reported $p$-values are anticonservative by orders of
magnitude, and only the effect sizes should be read as substantive. A ZIP-code-level analysis with
an explicit spatial error structure is the right tool and we have not carried it out.

A more basic point subsumes this one. We observe every California ZIP code in the range 90001--96162 ---
$1{,}705$ of them on our common universe --- and each one is labelled underserved or not according
to whether its boundary meets one of the nineteen triangles, a question settled entirely by its
position on the map. There is no sample and no random assignment, so there is no larger population
from which the observed association could be a chance departure, and the $p$-values do not have
their usual meaning. This is why we report Cram\'er's $V$ and the raw shares alongside every test,
and why \S\ref{sec:results-sensitivity} substitutes a sensitivity analysis for the additional
significance testing one might otherwise reach for. 
 
\subsection{Service Accessibility Limitations}

It is important to note that while many DMV locations accommodate walk-in customers, not all services are available to walk-in clients at every location. This service differentiation means that analyzing walk-in accessibility would only capture a subset of available DMV services, potentially misrepresenting the overall accessibility of comprehensive DMV services across different locations. The walk-in filtration does sit at a higher overall scale than the appointment filtration, since the median walk-in wait of $63.0$ minutes is more than three times the appointment median of $19.0$; but the median pairwise dissimilarity of $250.8$ minutes still dominates both, so the walk-in landscape is not flat and is not driven entirely by wait times. By focusing on appointment-based services, our primary analysis captures the full range of DMV offerings; \S\ref{sec:results-sensitivity} reports what changes when the walk-in weights are used instead.

\subsection{Future work}\label{sec:future}

Several directions follow directly. Repeating the analysis with public transit rather than driving
times would speak far more directly to the equity question, and requires no new data. Comparing the
weighted filtration against an unweighted one would quantify how much of the structure is driven by
wait times rather than geography. Restricting the demographic comparison to the most persistent
features, rather than to all of them, would test whether topologically significant gaps carry
localized information that statewide aggregation dilutes. Real-time and seasonal wait time data
would allow a dynamic version of the analysis.
Building the population-side filtration sketched at the end of \S\ref{sec:discussion} --- vertices
at population centroids, vertex weights the travel time to the nearest office --- would address the
central limitation of this work rather than merely documenting it. A ZIP-code-level logistic model
of coverage status on median income, ethnic composition, and population density, with a spatial
error structure, would put the demographic comparison on the right unit of analysis.
Finally, the framework applies to other kinds of
resource coverage; food deserts are a natural next target.

\section{Acknowledgments}
This material is based on work supported by the National Science Foundation under Grant No. DMS-2149642 and the Sloan Grant under Grant No. G-2024-22394 while the authors participated in a program hosted by the Simons Laufer Mathematical Sciences Institute (formerly Mathematical Sciences Research Institute) in Berkeley, California, during the summer of 2025.

\printbibliography

\end{document}